# Field-Driven Evolution of Chiral Spin Textures in Thin Nanodisk of the Helimagnets

*Haifeng Du, Wei Ning, Mingliang Tian,\* and Yuheng Zhang*

*High Magnetic Field Laboratory, Chinese Academy of Science, Hefei 230031, Anhui, P. R. China and Hefei National Laboratory for Physical Science at The Microscale, University of Science and Technology of China, Hefei 230026, People's Republic of China*

ABSTRACT

The magnetic field-driven evolution of chiral spin textures in thin helimagnet nanodisk with varied size are investigated by means of Monte Carlo simulation. It is demonstrated that the complex spin texture may simply be regarded as the superposition of the edged state with in plane spin orientation perpendicular or parallel to the edge and the bulk state with the features similar to two-dimensional chiral magnetic films. With the increase of the external field, the proportion of the parallel spins of the edge state increases, and the spin textures finally transfers into edged magnetic vortex. The arrangement of skyrmions strongly depends on the disk size. In addition, the uniaxial anisotropy and dipolar coupling in certain ranges are able to stabilize a special magnetic vortex with Skyrmionic core while the disk size is comparable with the wavelength of helix state.

Recently, Skyrmions [1], a topological stable nano-sized magnetic vortex, has been proved to exist in chiral magnets including specific metallic alloys with B20 structure such as MnSi [2,3], FeGe [4,5], $Fe_xCo_{1-x}Si$ [6], two-dimensional Fe/Ir(111) film [7], and even insulating oxide $Cu_2OSeO_3$ [8]. In metallic helimagnet this topologically stable spin texture is able to efficiently couple with the spin current, and thus leads to the topological Hall effects [9-11] and strong spin-transfer torque effect [12,13]. In insulating one, Skyrmion can induce electric polarization, indicating the potential of the manipulation of the Skyrmions by an external electric field [7,14]. These discoveries not only opened up the possibility of using skyrmions as memory bits with low energy consumption for future spintronic devices [15], but also initiated extensive investigations in physics both experimentally and theoretically [16-19].

However, in bulk materials skyrmions survive only in a very narrow temperature-field (T-H) window [2,20]. By contrary, the stabilized skyrmions may be realized by reducing the dimension of materials from bulk to two dimensional (2D) [3,4,6]. These results naturally lead to a question whether the stabilization of skymions can be further enhanced by continuingly reducing the sample size from a 2D film to nano-elements, such as nano-disk, ring, or stripe. It was known that the magnetic vortex is the stable ground state in micro-sized soft magnetic disk in order to reduce the magnetostatic energy [21]. In analogous to the soft magnetic vortex, skymions are anticipated to be more stabilized in high symmetry nanostructured magnetic objects due to the magnetostatic couplings, such as nanodisk. Meanwhile, a number of very interesting and potentially useful memory and logic devices have been recently proposed through the formation and controlled manipulations of the spin textures in magnetic nano-elements [22,23]. Unfortunately, the study of skyrmions at nano-scale is lacking and their magnetic properties still remain unclear.

In the present paper, we calculate the equilibrium magnetization process of the helimagnets with uniaxial anisotropy and dipolar coupling in thin nanodisk by means of Monte Carlo simulation. The simulated results showed that the complex spin texture may be simply regarded as the superposition of edge state with their orientation of in plane spin perpendicular or parallel to the edge and bulk state with the features analogous to 2D chiral magnetic films. With the increase of the external magnetic field, the number of the parallel spins of the edge state increases, and finally the spin texture transfers into edged magnetic vortex. For bulk states, the field-driven spin textures proceed, in turn, from the helix state to countable skyrmions and the final ferromagnetic state. The arrangement of the skyrmions strongly depends on the disk size. In addition, for certain uniaxial anisotropy and dipolar coupling ranges, a special magnetic vortex with skyrmionic core is spontaneously formed even without the help of thermal fluctuation or external field when the disk size is comparable with the wavelength of helix state.

For this study, a generally accepted simplest Hamiltonian for thin disk of chiral magnets with Dzyaloshinsky-Moriya $(DM)$ interaction and uniaxial anisotropy is written as [1]

$$w = A(\nabla \boldsymbol{m})^2 - K_u m_z^2 - \frac{1}{2}\boldsymbol{m} \cdot \boldsymbol{h}_m + D\boldsymbol{m} \bullet (\nabla \times \boldsymbol{m}) \qquad (1)$$

where $\boldsymbol{m}$, A, $h_m$ and $K$ are respectively the unit vector of magnetization, ferromagnetic exchange constant, ,the stray field and the effective anisotropy constant with the easy axis perpendicular to the disk plane. For the Monte Carlo $(MC)$ simulation, we divided the disk into discrete blocks with unit magnetization $\hat{S}$, then, a 2D square lattice

Hamiltionian is written as [24-25]

$$E = -J\sum_{i<j}\hat{S}_i \cdot \hat{S}_j - \sum_{i,j} D_{R_{ij}} \cdot (\hat{S}_i \times \hat{S}_j) - K\sum_i (\hat{S}_i \cdot \hat{e}_i)^2$$
$$+ d\sum_{i<j}\left(\frac{\hat{S}_i \cdot \hat{S}_j}{|R_{ij}|^3} - \frac{3(\hat{S}_i \cdot \hat{R}_{ij})(\hat{S}_j \cdot \hat{R}_{ij})}{|R_{ij}|^5}\right) - H \cdot \sum_i \hat{S}_i \quad (2)$$

In sequence, where the five terms denote the ferromagnetic exchange coupling with exchange constant $J$, the $DM$ interaction with constant $|D_{R_{ij}}| = D$ pointing the vector along site $i$ and $j$, the uniaxial anisotropy energy with constants $K$, the dipolar-dipolar interaction between the blocks with dipolar strength constant $d$, and the Zeeman energy with external field perpendicular to the disk plane, respectively. The ration $D/J$ with $D=1$ was chosen to yield the spiral propagation wavelengths of $T_s = 10/\sqrt{2}$ lattice constants. Since in the discrete model both direct and $DM$-exchange in an inhomogeneous state will have slightly different energy, depending on their orientation with respect to the underlying discrete bond orientation of the lattice, the propagation direction of the helix state was locked into the $\langle 11 \rangle$ directions even without the weak crystal field energy [6,24], which determined the propagation direction of the helix state in the continuous model. A detailed investigation into the effect of crystalline field interactions on the spin arrangement in two-dimensional film of the chiral magnets has been performed by Yi *et al* [24]. Here, we neglected this weak energy, which is also partially equivalent to the anisotropy arising from the discretization of continuous model. A high temperature annealing metropolis algorithm is used to obtain the equilibrium spin configurations [6]. At each temperature the system is allowed to relax towards equilibrium for the first $10^5$ Monte Carlo steps and thermal averages are calculated over the subsequent $10^5$ steps. After obtaining the final equilibrium

magnetic configurations, the relative local chirality $\tau_r$ at lattice $r$ is calculated by [24]

$$\tau_r = \hat{S}_r \cdot (\hat{S}_{r+\hat{x}} \times \hat{S}_{r+\hat{y}}) + \hat{S}_r \cdot (\hat{S}_{r-\hat{x}} \times \hat{S}_{r-\hat{y}}) \qquad (3)$$

The total chirality $\tau$ of the disk is calculated by $\tau = a \sum_r \tau_r$ with the constant $a$, which is chosen to grantee the unit topological charge of one skyrmion.

Firstly, let's only consider the main energy scales, i.e. direct and $DM$-exchange couplings, in Eq.(2). A typical field-driven evolution of the chiral modulations for the disk with radius $R_d = 10$ is shown in Fig.1. Discontinuous jumps in $\tau(H)$ and $m_z(H)$ curves mark the system with different spin textures. In bulk materials, the theoretical calculation has proved that only two first order transition lines exist, named as $H_h$ and $H_s$, corresponding to the evolutions from helix state to skyrmion lattice and subsequently to ferromagnetic state, respectively [1]. In the present system, the situation is more complex. For $H < H_h$ with the critical field $H_h = 0.2$, it is clearly observed that the helix state is distorted in the edge, favoring their in-plane spin orientation perpendicular or parallel to the edge, as illustrated in the Fig.1.(b). A similar phenomenon has been observed in the stripe domain of micro-sized Ni wires with large perpendicular anisotropy [26,27]. The distortion of spin textures in the edge results in non-zero chirality. In rigorous continuous model Eq.(1) just including direct and $DM$-exchange interactions, the distorted helix state is expected to be isotropy and continuously degenerate with respect to propagation directions in nanodisk. However, as it expounded in the previous section, the discretization of the isotropic model in the $MC$ simulation fixed the propagation direction of the helix state into the $\langle 11 \rangle$ directions, leading to fourfold degenerate ground states with every in plane $90°$ rotation. In addition, two types of spin configurations with different symmetry are observed. For $H < 0.1$, in-plane spin $m_{xy}$ shows mirror symmetry and $m_z$ displays two-fold rotational symmetry with respect on the $<11>$ axis, as displayed in Fig.1.(b). If $H > 0.1$, the symmetry of $m_z$ and $m_{xy}$ exchanged each other, as illustrated in the inset of Fig.1.(a) $(H = 0.18)$.

For $H_h \leq H < H_{S1}$ with the critical field $H_{S1} = 0.29$, The system undergoes a intermediate state before entering skyrmion state. With the increase of $H$, the proportion of the parallel spins of the edge state increases, and the spin texture finally transfers into edged magnetic vortex. The typical intermediate state in Fig.1.(c) is characterized by two compact bimerons surrounded by the edged vortex. Indeed, the edged vortex, favoring its in-plane spin orientation completely parallel to the edge, is related to the rotational symmetric solutions of Euler equations of the helimagnets. The detailed numerical analysis on the edged vortex state has been investigated in Ref.[28]. The bimeron, composed of two half-disk domains with nonvanishing $1/2$-topological charge and a rectangular stripe domain with zero topological charge, has been predicted to exist in two-dimensional easy-plane ferromagnets with $DM$ interactions at finite temperature [29]. Here, the countable meron provides a good opportunity to investigate the properties of isolated bimeron.

For $H_{S1} \leq H < H_E$ with the critical field $H_E = 0.77$ that defines the transition line from skyrmions to edged vortex, skyrmions have countable number. Both skyrmions in the interior of disk and the edged vortex follow the same sense of the rotation determined by the sign of $D$ [1], but the spins in the center of the two spin textures point to two opposite directions, resulting in opposite rotation directions of in-plane spins. According to the Eq.(3), the total chirality of the system is lower than that of pure skyrmions. To account for the transformation in details, we further divided this region into two parts separated by the transition line

$H_{S2} = 0.55$, i.e. the region with maximal skyrmion numbers is called skyrmions, and the rest is named as skyrmion gas. Finally, for $H > H_E$ only the edged vortex exists, as displayed in Fig.1.(d). Numerical calculation has proved that Eq.(1) includes axisymmetric solutions with certain boundary conditions [30-32]. In the case of disk, neglecting the edge pinning effect, the global minimal energy corresponds to the natural boundary conditions [28], which results in the nonvanishing in-plane projects of spins at the edge even at high field. This was clearly observed by non saturation magnetization $m_z$, i.e., the edged vortex is protected by the naturally boundary conditions.

In magnetic nano-elements, the sample size always plays a vital role. In Fig.2.(a), we plot the normalized $\tau(H,R)$ as the functions of external field and disk size. Some representative snapshots of $m_z$ distribution are shown in Fig.2.(b). For $R \leq R_c$ with the critical size $R_c \approx 5 = 0.8T_h$, other than skyrmion but incomplete helix state with "$U$" shape (Fig.2.(b) $R = 5, H = 0.2$) formed. This result has an important implication for the future experiments by setting a limit on the skyrmion-based nanodevice. For $6 \leq R \leq 8$, the transformations from helix state to skyrmions with the maximal skyrmion numbers $N_S^m = 1$ (Fig.2.(b) $R = 8, H = 0.4$) and subsequently to edged vortex are all first order. For $9 \leq R \leq 12$, the evolution of spin textures in the external field is similar to that of $R = 10$, where the intermediate state and skyrmions gas arise. Especially, for $R = 9$, different initial simulated parameters bring out different spin textures with nearly identical energy under the same field (Fig.2.(b) $R = 9, H = 0.38$). These data indicated that the system may be degeneration with different skyrmion numbers at some certain fields. For $13 \leq R \leq 17$, skyrmion arrangement is characterized by one in the center surrounded by the rests. For

$R > 17$, $<\tau(H)>$ curves gradually transferred into continuous ones. Instead of forming ideal hexagonal skyrmion crystals observed in bulk or two-dimensional film, the skyrmions shift about like those of a liquid, but a freeze frame would reveal that the system has long-range order like a solid. Each skyrmion in the interior of disk has on average six neighbors, forming the hexatic phase [33].

To get a general insight into the influence of the uniaxial anisotropy and dipolar-dipolar interaction on the magnetic structure, we extended the model into the most common situation by varying $d$ and $K$ for fixed $D$ and $J$. The resulting magnetic phase diagram without the external field in Fig.3.(a) for $R = 7$ is composed of four regions: (Ⅰ) distorted helix state, (Ⅱ) chiral soft magnetic vortex for stronger dipolar couplings, (Ⅲ) chiral stripe domain for stronger uniaxial anisotropy, and (Ⅳ) special magnetic vortex with skyrmion core. The formation of spin textures in region (Ⅱ) and (Ⅳ) is easily understood as the counterparts of the vortex in microsized soft magnetic nanodisk [22] and stripe domain in two dimensional films with high uniaxial anisotropy [27], respectively. Antisymmetric $DM$ couplings lift the degeneracy and put the chiral feature into the system. Recent experiments provide clear evidence for the chiral magnetic vortex in $FeNi$ alloy [34] and chiral stripe domain in $Fe(2ML)/Ni/Cu(001)$ films [35]. In region (Ⅳ), the special vortex with skyrmion core has also been analyzed in detail by numerical methods in Ref.[28], where the special vortex corresponds to the axis symmetry solution of Eular Equation of Eq.(1) without magnetostatic energy. In particular, for some certain disk size, the energy density of the special vortex is lower than that of helix state in 2D films of chiral magnets, indicating a big chance to form spontaneous vortex ground state even without the help of uniaxial distortions and dipolar couplings. Here, the formation of this vortex ground state required the combination of the uniaxial anisotropy and dipolar couplings. The discrepancy between numerical method and Monte Carlo simulation probably results from the discretization of Eq.(1), leading to some inaccuracy of Monte Carlo simulations [36]. Indeed, this special vortex has partially been observed in FeGe films by Lorentz microscopy, where incomplete circle helix state formed in the irregular inhomogeneous sample [37]. More recently, the homogenous thickness-controllable FeGe films with skyrmion state have been fabricated by magnetron sputter [38]. These advances provide a good basis to fabricate nanodisk of helimagnets, and then to realize the special vortex ground state.

The effect of $d$ and $K$ on the equilibrium magnetization process of the disk is displayed in Fig.3(b) and 3(c). For $d=0$ with increasing $K$ up to $1.0$, both $H_h$ and $H_E$ decrease, analogue to the theoretical results in 2D films of chiral magnets [39]. For $K>1.0$, only edged vortex exists in spite of the amplitude of magnetic field. For $K=0$ with increasing $d$ up to the critical value $0.2$, above which chiral soft magnetic vortex formed in Fig.3 (a), $H_h$ almost keeps unchanged, while $H_E$ increases linearly. These features are significantly different from those in 2D films of the chiral magnets where both $H_h$ and $H_E$ increase linearly with increasing saturation magnetization [1], indicating a high stability for skyrmions in the nandisk of helimagnets. For $K=0$, $d>0.2$, corresponding to the region (Ⅱ) in Fig.3.(a), the existing interval of magnetic field for skyrmions decreases with increasing $d$. Compared with the pure soft magnetic vortex where only vortex state exists in the whole magnetization process, moderated $DM$ couplings are able to produce skyrmions in the soft magnetic disk under certain intervals of the magnetic field. More importantly, in region (IV), because the ground state is skyrmions, only one transition from skyrmions to edged vortex is observed with increasing $H$, as shown in Fig.3.(c). Moreover, the maximal number of skyrmions $N_S^m$ is raised to 2 in contrast to $N_S^m =1$ for the case without $K$ and $d$. The similar magnetic phase diagram with different $R$ is plotted in Fig.3 (d) ($R=10$) and 3 (e) ($R=15$) It was found that the region (IV) shrinks with the increase of $R$. When $R=20$, the region (IV) disappears completely (not shown). Besides, the phase transition from region (IV) to (II) or (I) to (II) is second order, where chiral multi-vortex magnetic domains are observed in the region of transit phase [40].

It must be noted that all the present discussion is only involved to the equilibrium state. In real materials, magnetization process depends on the magnetization history due to the existence of metastable state. In addition, the use of a cubic mesh might also result in somewhat uncertainty or inaccuracy due to the artificial edge roughness and discretization of Eq.(1). Even so, we do believe that the simulated results may grasp some main properties of the nandisk of helimagnets.

In conclusion, we calculate the equilibrium magnetization process of the helimagnets with uniaxial anisotropy and dipolar coupling in thin nanodisk by means of Monte Carlo simulation. The complex spin textures may simply be regarded as the superposition of edged state with their orientation of in plane spin perpendicular or parallel to the edge and bulk state with the features analogous to 2D chiral magnetic films. With the increase of the external field, the proportion of the parallel spins of the edge state increases, and finally the spin texture transfers into edged magnetic vortex, which is protected by the naturally boundary condition. The arrangement of skyrmions strongly depend on the disk size, in some aspect, similar to the vortex lattice in a confined Type-Ⅱ superconductor [41]. In addition, For certain uniaxial anisotropy and dipolar coupling ranges, a special magnetic vortex with skyrmionic core is spontaneously formed while the disk size is comparable with the wavelength of helix state.

**ACKNOWLEDGEMENTS**: This work was supported by the National Key Basic Research of China, under Grant Nos. 2011CBA00111 and 2010CB923403; the National Nature Science Foundation of China, Grant No. 11174292 and No. 11104281，No. 11104280;

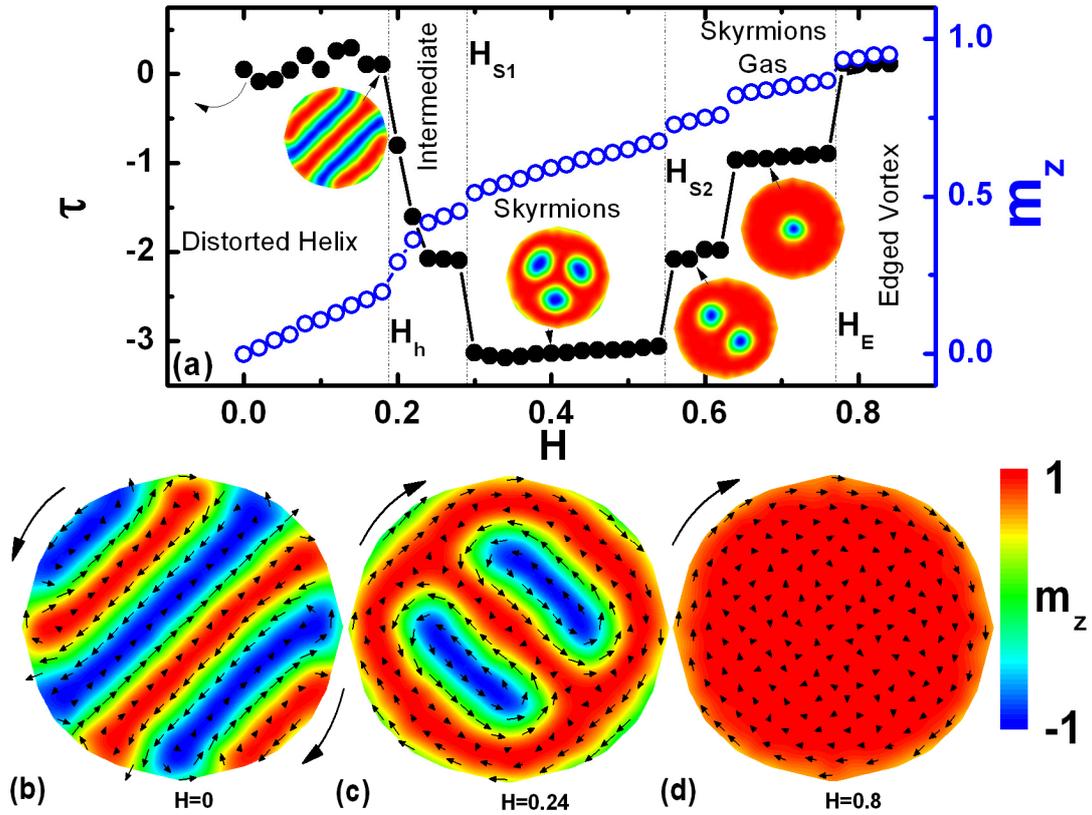

Fig.1. (a) The normalized magnetization $m_z$ and chirality $\tau$ of disk for $R=10$ as the function of magnetic field $H$, discontinuous curves $m_z(H)$ and $\tau(H)$ indicating different spin textures, the corresponding snapshots of out of plane magnetization $m_z$ is shown in the inset; The full spin orientation for three typical magnetic field (a): for zero field $H=0$, the complex spin textures may be regarded as the superposition of the edged state with in plane spin orientation perpendicular or parallel to the edge and bulk state with the features of helix. (b): for moderate field $H=0.24$, the edged state and helix state transferred into edged vortex and bimeron respectively; (c): for high field $H=0.8$, the Skyrmions in the center of disk transferred into ferromagnetic state, while the edged vortex still exists.

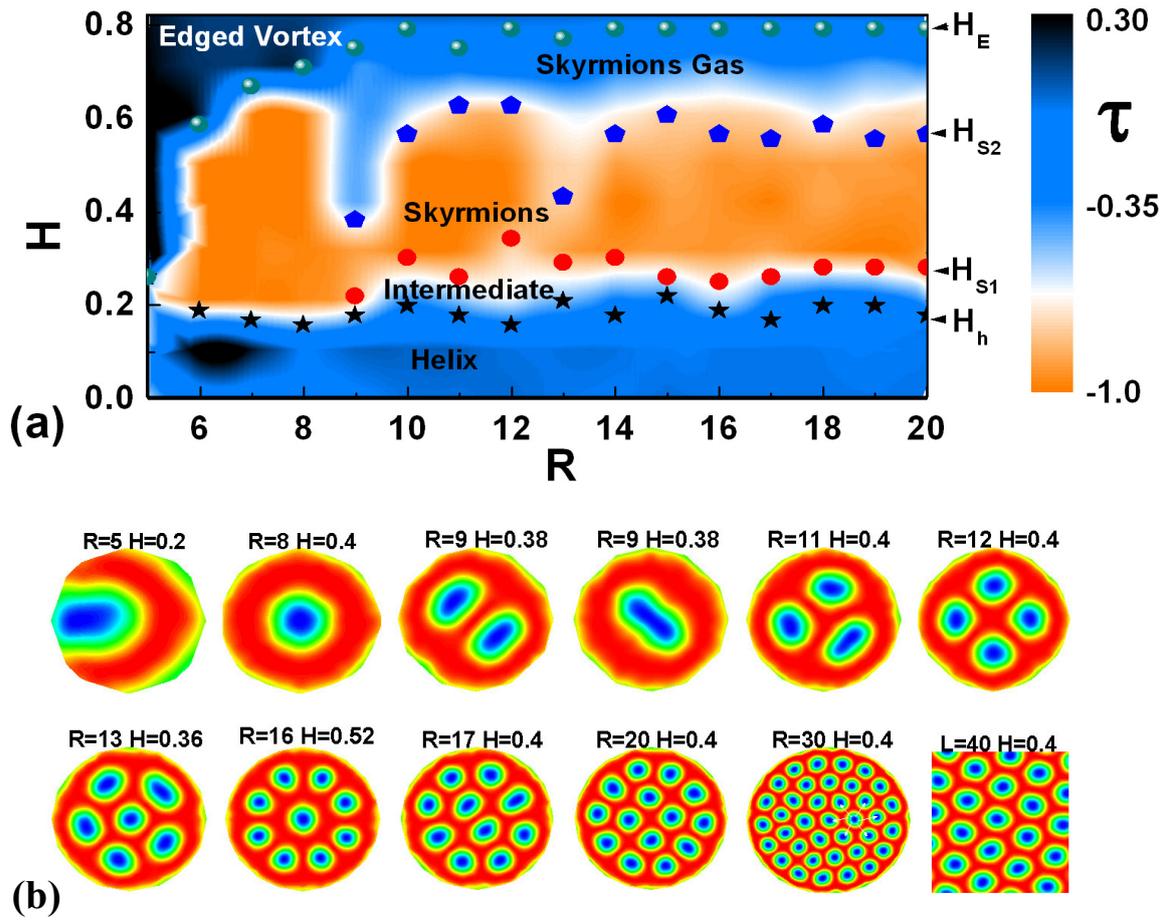

Fig.2. (a): The phase diagram as the functions of external field $H$ and disk size $R$ based on the normalized curves $\tau(H,R)$; (b): The dependence of the arrangements of Skrymions on the disk size, while $R \leq 5$, no Skyrmions formed. With increasing $R$, the Skyrmions number increases discontinuously and sensitively depends on the disk size. For comparison, the spin arrangement of Skyrmions in two-dimenional films is also displayed $(L=40, H=0.4)$.

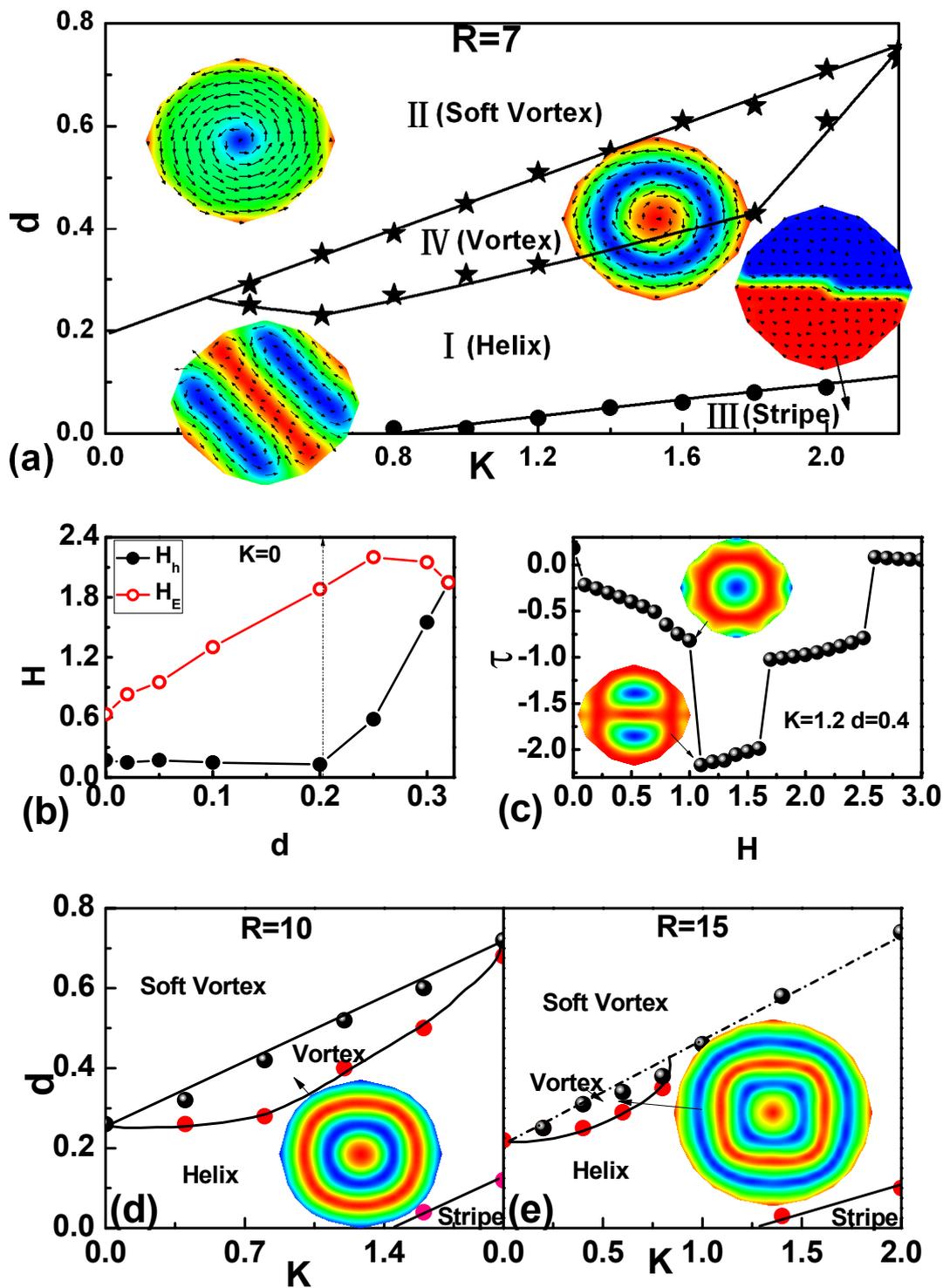

Fig.3.(a): The magnetic phase diagram as the functions of $d$ and $K$ for fixed $D$ and $J$ and $R = 7$, it is composed of four regions: (Ⅰ) distorted helix state, (Ⅱ) chiral soft magnetic vortex for stronger dipolar couplings, (Ⅲ) chiral stripe domain for stronger uniaxial anisotropy, and (Ⅳ) special magnetic vortex with skyrmions core. The corresponding snapshots of out of plane are illustrated in the inset. (b):

The dependence of existing magnetic intervals of Skyrmions on the dipolar couplings constant $d$. (c): Typical evolution of spin textures in region (IV). The magnetic phase diagram as the functions of $d$ and $K$ for fixed $D$ and $J$ for (d): $R=10$ and (e): $R=15$, with increasing $R$, the region (IV) shrinks.